\documentclass[10pt]{revtex4}
\usepackage{graphicx}
\usepackage{amsmath,amssymb}
\usepackage{mathtools,physics}

\begin{document}

\title{Interference in the Heisenberg Picture of Quantum Field Theory, \\Local Elements of Reality and Fermions}
\author{Chiara Marletto, Nicetu Tibau Vidal and Vlatko Vedral}
\affiliation{Clarendon Laboratory, University of Oxford, Parks Road, Oxford OX1 3PU, United Kingdom\\Centre for Quantum Technologies, National University of Singapore, 3 Science Drive 2, Singapore 117543\\
Department of Physics, National University of Singapore, 2 Science Drive 3, Singapore 117542}

\begin{abstract}
We describe the quantum interference of a single photon in the Mach-Zehnder interferometer using the Heisenberg picture. Our purpose is to show that the description is local just like in the case of the classical electromagnetic field, the only difference being that the electric and the magnetic fields are, in the quantum case, operators (quantum observables). We then consider a single-electron Mach-Zehnder interferometer and explain what the appropriate Heisenberg picture treatment is in this case. Interestingly, the parity superselection rule forces us to treat the electron differently to the photon. A model using only local quantum observables of different fermionic modes, such as the current operator, is nevertheless still viable to describe phase acquisition. We discuss how to extend this local analysis to coupled fermionic and bosonic fields within the same local formalism of quantum electrodynamics as formulated in the Heisenberg picture. 
\end{abstract}

\maketitle

\section{Introduction}

The superposition principle of quantum theory has an immediate expression in the Schr\"odinger picture. If two different state vectors satisfy the Schr\"odinger equation then their linear sum also does. Thus the Schr\"odinger picture lends itself naturally to the description of quantum interference. For example, if the state $|0\rangle$ of a qubit acquires a phase $\phi_0$ and the state $|1\rangle$ acquires a phase  $\phi_1$, then according to the Schr\"odinger equation the superposed state of the qubit evolves into $e^{i\phi_0}|0\rangle + e^{i\phi_1}|1\rangle$. This phase difference can then be detected by performing a measurement in the $|\pm\rangle = \frac{1}{\sqrt{2}}(|0\rangle \pm |1\rangle)$ basis. \\

By contrast, in the Heisenberg picture, the initial state vector (the so-called Heisenberg state) never changes. It can always be assumed to be a fixed state $\rho_0$ without any loss of generality. How then is the interference phenomenon manifested? 
The explanation of interference lies in the dynamics of operators which evolve unitarily, consistently with the dynamical evolution of the state vectors in the Schr\"odinger picture.  The consistency condition is given by the requirement that the Heisenberg and Schr\"odinger pictures must be empirically equivalent. For any given observable $\hat O$ of a physical system, the empirically accessible quantity at any one time is given by the expected value $\Tr(  \hat O \rho (t) )=\Tr( \hat O(t) \rho_0 )$, where $\hat O(t)=U(t)^{\dagger} \hat O U(t)$ and $\rho(t)=U(t) \rho_0 U^{\dagger}(t)$. Hence the empirical content of the two pictures is the same.  But the mode of explanation is different, especially regarding locality \cite{Deutsch}.\\ 

Here we discuss the advantages of explaining interference (the spatial one, in particular) in the Heisenberg picture of quantum field theory. 
First, we illustrate how the Heisenberg picture provides local elements of reality in terms of {\sl q-numbered descriptors} (quantum observables) for bosonic and fermionic fields, both free and interacting, much as in classical field theory. However, the local elements of reality are in the quantum case operators (q-numbers) and not c-numbers (the usual complex numbers, that all commute with each other).\\

These q-numbered {\sl descriptors} satisfy the {\sl principle of no-action at a distance}: given a partition of the whole universe into subsystems, operations (unitaries or CP-maps in general) involving the descriptors of a given subsystem cannot modify the descriptors of other non-overlapping subsystems. In \cite{Deutsch} the proof of this fact is presented for an $N$-qubit system and extended to any other physical system via the universality of quantum computation. However, there are some outstanding points to clarify in the case of quantum fields. For instance, fermionic fields lack crucial qubit properties (such as local tomography \cite{Ariano}); hence the proof of locality by the universality of computation may not directly apply to them. \\

\section{Bosonic Mach-Zehnder interferometry}

Consider a concrete example of bosonic interferometry. A single-photon going through a Mach-Zehnder interferometer has been the foremost way of thinking about interference in quantum information, and computation \cite{Nielsen}. It is analogous to the double-slit experiment, which in the words of Feynman contains ``the only mystery" in quantum physics \cite{Feynman}. We shall imagine that an additional phase $\phi$ is introduced locally in one of the arms (how it is applied depends on the physics of the system undergoing interference and is irrelevant for our discussion). \\

We consider the standard quantisation procedures of the electromagnetic field, which lead to introducing the creation and annihilation bosonic operators at mode $x$, denoted by $a_x, a_x^{\dagger}$. These operators satisfy the following constraints:  $[a_x, a_y]=0$, $[a_x,a_y^\dagger]=\delta_{x,y}$, and $a_x\ket{0}=0$, where $\ket{0}$ is the chosen vacuum state of the global Fock space and $[A,B]=AB-BA$ is the commutator of the two operators $A$ and $B$. 
For simplicity of exposition, we shall assume that $x$ represents a region of space where the photon can be confined to arbitrarily high accuracy; in this case, $x$ can be either $L$ (a region around the left arm of the interferometer) or $R$ (a region around the right arm of the interferometer). These two regions are non-overlapping (their separation being much larger than their respective extents). The details of the free evolution of the photons are not relevant for present purposes. Also, we do not consider the polarisation of the photons, since it is irrelevant for the locality discussion.  \\ 

In the Schr\"odinger picture, the quantum state of the photon changes after the first beam splitter ($U_{BS}$), then acquires the additional phase in one arm (say the left one) via $U_{L}^{(\phi)}$, and finally undergoes another change at the final beamsplitter ($U^\dagger_{BS}$). Labelling the quantum state where the photon is on the left or right arm of the interferometer respectively as $\ket{L}\doteq a_L^{\dagger}\ket{0}$ and $\ket{R}\doteq a_R^{\dagger}\ket{0}$, the dynamical evolution of the photon is given by:

\begin{eqnarray}
|L\rangle \xrightarrow[\text{}]{U_{BS}} \frac{1}{\sqrt{2}}(\ket{L}+\ket{R}) \xrightarrow[\text{}]{U_{L}^{(\phi)}}  \frac{1}{\sqrt{2}} (|R\rangle + e^{i\phi} |L\rangle) \xrightarrow[\text{}]{U_{BS}}  \frac{1}{\sqrt{2}}(|+\rangle + e^{i\phi} |-\rangle) =\sin \phi/2 |L\rangle + \cos \phi/2 |R\rangle \; ,
\end{eqnarray}
where $\ket{\pm}=\frac{1}{\sqrt{2}}(\ket{L}\pm\ket{R})$ are equally weighted superpositions of the left and right path.\\

In the Heisenberg picture, we see how the phase affects the dynamical evolution of photonic quantum observables. We can consider for instance the operator representing the vector potential field of each position mode $x$, which is proportional to $A_x=a_x+a_x^{\dagger}$. (The same analysis could be done with the electric or magnetic field). Since we have two arms of the interferometer, $x=L$ (left) and $x=R$ (right), we shall specify the field in both of these (which we think of as modes of the electromagnetic field). We will use the ordered pair notation $t: [\![a_L(t)+a_L(t)^{\dagger}\;;\; a_R(t)+a_R(t)^{\dagger}]\!]$ to denote the descriptors of the left and right modes at time $t$, where the left mode descriptor occupies the first slot of the ordered pair and right mode occupies the second slot. (Note that in general, due to unitarity, it is possible to retrieve the dynamical evolution of {\sl all} relevant observables of a composite system by merely tracking the dynamical evolution of the generators of the algebra of observables of each subsystem, \cite{Gottesman}. In this case, the generators would be $a_L(t), a_R(t)$.) 

At the start, let the field operators be 
\begin{equation*}
    t_0: [\![ a_L+a_L^{\dagger} \;;\;  a_R+a_R^{\dagger} ]\!]\;
\end{equation*}
 and the Heisenberg state $|\Psi\rangle = |1_L 0_R\rangle$. This simply represents the quantum photon field operators in the left and the right modes at time $t_0$, with one photon existing in mode $L$. The unitary beam splitter $U_{BS}$ applied at time $t$ acts as Bogoliubov transformations on the creation and annihilation operators: $a_L(t) \xrightarrow[\text{}]{U_{BS}(t)} a_L(t+dt)=\frac{1}{\sqrt{2}}(a_L(t) + a_R(t))$ and  $a_R(t) \xrightarrow[\text{}]{U_{BS}(t)} a_R(t+dt)=\frac{1}{\sqrt{2}}(a_L(t) - a_R(t))$. \\
 
 So the photon field operator descriptors after the first beam splitter, expressed as functions of the initial descriptors, are: 
 \begin{equation}
 t_1: [\![\frac{1}{\sqrt{2}}(a_L + a_R+a_L^{\dagger}+a_R^{\dagger}) \;;\;  \frac{1}{\sqrt{2}}(a_L - a_R+a_L^{\dagger}-a_R^{\dagger})]\!]\;    
 \end{equation}

The phase shift $U_{L}^{(\phi)}(t)$ acts only on the left arm. That is, it is a function of the operators $a_{L}(t)$ only. Hence, it induces a change in the left modes only: $a_L(t) \xrightarrow[\text{}]{U_{L}^{(\phi)}(t)} a_L(t+dt)=e^{i\phi}a_L(t)$ and $a_R(t) \xrightarrow[\text{}]{U_L^{(\phi)}(t)} a_R(t+dt)=a_R(t)$. The new field operators after the phase shift are therefore:
\begin{equation}
    t_2: [\![ \frac{1}{\sqrt{2}}(e^{i\phi}(a_L + a_R)+e^{-i\phi}(a_L^{\dagger}+a_R^{\dagger})) \;;\;  \frac{1}{\sqrt{2}}(a_L- a_R+a_L^{\dagger}-a_R^{\dagger})]\!]\;
\end{equation}

The property of no-action at a distance is the crux of quantum field theory in the Heisenberg picture: changes induced by a phase shift acting locally on one mode do not affect operators pertaining to the other modes. In our example, only the field operators of the left mode (the first slot of the ordered pair above) contain the phase, while the field operators of the right mode do not. Hence by simple inspection of the descriptors of the two modes, we can tell where the phase shift was applied. Note also that a state-tomography of the left mode would not at this stage reveal the phase (the expected value of the number operator of the left mode does not depend on the phase). Hence the phase is at this stage encoded in the left mode, but it is locally inaccessible.\\   

The final step is the second beam splitter, which induces once more a mixing of modes, and hence allows one to recover the phase. The field operators at the output of the interferometer are then:

\begin{equation}
   t_3:  [\![\cos(\frac{\phi}{2})a_L+i\sin(\frac{\phi}{2})a_R + {\rm h.c.} \;;\;  \cos(\frac{\phi}{2})a_R-i\sin(\frac{\phi}{2})a_L + {\rm h.c.}]\!]\;
\end{equation}

The interference is manifested when we take the expected value of the number operator $\hat{N}_x(t)=a^\dagger_x(t) a_x(t)$ at time $t_3$, in the Heisenberg state $|\Psi\rangle = |1_L 0_R\rangle$. This value can be calculated from $\langle A^2_{x}(t_3) \rangle_{\ket{\Psi}}=1+2\langle \hat{N}_x(t_3) \rangle_{\ket{\Psi}}$ . Thus the phase (that we have tracked with $A_x(t)$ is now manifested in a direct observation. For the output left mode, we obtain:
\begin{gather}
 \langle \hat{N}_L(t_3) \rangle_{\Psi}=\cos^2 \frac{\phi}{2} \; .
\end{gather}

The expected value of the output mode $R$ could be calculated in the same fashion (and it would yield the value of $\sin^2 \phi/2$). Hence the expected values at the end of the interferometry are the same in the Heisenberg and Schr\"odinger pictures, as they are empirically equivalent. The significant difference in the explanation for the interference is that in the Heisenberg picture the phase introduced by the phase shift on one mode is only manifested in that mode and not others, locally. In the Schr\"odinger picture this would not be the case as the wave-function does not allow for a separable description and the phase difference due to the beam splitter acting on mode $L$ could equally well have been introduced by a beam splitter acting on mode $R$.  In \cite{Deutsch} , this fact is pointed out in regard to the quantum teleportation protocol with qubits. The key advantage of the Heisenberg picture is that it is manifestly local (i.e., it satisfies no action at a distance as well as no-signalling; of course, the Schr\"odinger picture does not allow signalling either and, in this sense, it is also local, or micro-causal in the language of field theory).\\

Any bosonic field (say pertaining to a Bose condensate of atoms) has precisely the same description as above in the Heisenberg picture. We can interfere condensates applying this Mach-Zehnder interferometer implementation, and the operator description of this interference would be identical to the one presented above. The same is true of fermionic fields, but with one significant subtlety, due to the presence of superselection rules. To expose this subtlety, we will now proceed to describe a single electron Mach-Zehnder interferometer. 

\section{Fermionic Mach-Zehnder interferometry}

For a single electron, one could naively expect that the model for interferometry is the same as in the bosonic case, replacing the descriptors with fermionic creation and annihilation operators $f_x$ and $f_x^{\dagger}$ at mode $x$. This strategy, however, is not possible. Fermionic operators anticommute at space-like separated points: $\{f_x , f_y^{\dagger}\}=\delta_{x,y}$ and $\{f_x , f_y\}=0$, where the curly brackets represent the anticommutator, $\{A, B\}\doteq AB+BA$. Hence, the operator $f_x+f_x^{\dagger}$ is not an observable. If it were, then one would be able to signal. To explain why, we present a simple argument which is originally due to Wigner \cite{Wigner}. It leads us to have to impose a superselection rule (the parity superselection rule), ruling out superpositions of even- and odd-numbered fermionic states.\\

The thrust of the argument is that, if one could superpose even and odd numbers of fermions, such as by preparing an eigenstate of $f_x+f_x^{\dagger}$, one would violate the no-signalling principle. For instance, one could send messages between two space-like separated regions by the following protocol.\\

Consider two space-like separated regions, $A$ and $B$. Suppose that $f_A$ is a fermionic mode at $A$, and $f_B$ is one at $B$. First, let us consider the Schr\"odinger picture. Let us assume that $\frac{1}{\sqrt{2}}(|0\rangle+|1\rangle)_B$ (eigenstate of $f_B+f_B^{\dagger}$ with eigenvalue $1$) is an allowed fermionic state. Then, let us prepare the product state $|0\rangle_A (|0\rangle+|1\rangle)_B$. Assuming no restrictions on the allowed quantum observables, $i(f_A^{\dagger}-f_A)$ is an allowed Hamiltonian; hence the unitary $U_A=\exp(\frac{\pi}{2}(f_A^{\dagger}-f_A))$ is also permitted, which can create one fermion state out of the vacuum in the mode $A$. Now, let us either leave $A$'s state as the vacuum or create one electron in $A$, by applying $U_A$. If the state remains the vacuum, $B$'s state remains unchanged: the expected value of $f_B+f_B^{\dagger}$ is $+1$ at the end of the protocol. If an electron is created in the region $A$ by applying $U_A$, the state becomes $|1\rangle_A (|0\rangle+|1\rangle)_B$. Now, the expected value of the observable $f_B+f_B^{\dagger}$ at $B$ is $-1$ in this state. Thus an observer at $B$, if they were able to prepare and distinguish superpositions of the vacuum and one electron state, they could tell whether an electron had or had not been created at $A$, despite $A$ being space-like separated from $B$! Generalising the argument, it follows that any superposition of even and odd numbers of fermions is prohibited. \\

An equivalent argument exists in the Heisenberg picture.  Assume that the Heisenberg state is $\ket{\psi_H}=|0\rangle_A (|0\rangle+|1\rangle)_B$. We can see what happens to the observable $f_B+f_B^{\dagger}$ when acted upon by $U_A$;  corresponding to the action of creating a superposition of even and odd numbers of particles. One obtains: $U_A^\dagger(f_B+f_B^{\dagger})U_A=-(f_B+f_B^{\dagger})$. This expression is a violation of the principle of no-action at a distance because quantum observables at $B$'s location can be modified instantaneously by the action of a unitary $U_A$, which is only operating on $A$.  Taking the expected value of the field operators at B at the end of the protocol, $U_A^\dagger(f_B+f_B^{\dagger})U_A$, using the Heisenberg state $\ket{\psi_H}$, we obtain $-1$, reaching the same conclusion as in the Schr\"odinger picture.\\

Following this line of argument, one imposes a parity superselection rule: fermionic observables of a given mode $x$ must commute with the parity operator $\exp(-i\pi f_x^{\dagger}f_x)$; which implies that they have to consist of quadratic forms of fermionic operators, such as the electric charge density operator: $j_0(x)=-ef_x^{\dagger} f_x$. The quadratic forms of fermionic operators at different space-like points commute (just like in the bosonic case), so there is no problem with locality.\\ 

However, there is more. Quadratic forms of fermionic operators alone cannot keep track of the phase in the Mach-Zehnder experiment in the same way as bosonic operators can. So not only would the phase not be registered in the local density operator (or any other local quadratic observable ); even in the Heisenberg picture, it would not be registered in any local quadratic fermionic descriptor. Luckily, the second-quantised Dirac field does not contain just the electron operators; it also contains the positron operators. So in the Heisenberg picture, it is still possible to track local observables of the Dirac field pertaining to each mode, in order to give an entirely local account of the interferometry.\\ 

In the bosonic case, any unitary process can be fully described by tracking observables, since the algebra generators can be found from observables by $a_x=(a_x+a_x^\dagger)-i(i(a_x-a_x^\dagger))$. However, for fermions, this is not entirely the case (in \cite{Nicetu} a detailed local mathematical analysis of general fermionic systems that goes beyond the scope of the current paper shall be presented).\\ 

Going back to the fermionic Mach-Zehnder interferometer, the proper second-quantised Dirac field is described by the four-spinor operator (see \cite{Weinberg})
\begin{equation}
\psi (x) = b(x) + d^{\dagger}(x)\;.
\end{equation}

This involves the electron annihilation operator $b(x)$ and the creation operator of a positron $d^\dagger(x)$ at point $x$. 

Once more, here we deliberately omit the spinor details and the momentum representation as it is not relevant for the following argument -- for details see e.g.\cite{Schweber}. This fermionic Dirac field descriptor is not Hermitian and thus is not an observable. Also, the superselection rules prohibit superposing odd and even numbers of fermions. Thus, no linear combination of creation and annihilation operators is allowed to represent a physical variable.\\ 

To describe the Mach-Zehnder electronic experiment we need to construct a quadratic operator out of the fundamental descriptor of the Dirac field. Consider for instance the charge density operator pertaining to each arm mode $(j_0(x_L),j_0(x_R))$, where $j_0(x)=-e:\psi(x)^{\dagger} \psi(x):=-e(b^\dagger_x b_x -b_x d_x +b^\dagger_x d^\dagger_x-d^\dagger_x d_x)$ (we would in general have to use the 4-vector including also the current density, but, in this case, the other 3 components do not add to our analysis). We denote by $:AB:$ the normal ordering of the fermionic operators $A$ and $B$. Even though $d_x,b_x$ are spinors with some orthogonality properties imposed, all the four terms are non-zero in general. We will also assume, for simplicity of exposition, that the Heisenberg state is $|\Psi\rangle_{e p} = \frac{1}{\sqrt{2}}(|0_L 1_R\rangle_e + |1_L 0_R\rangle_e)|0_L0_R\rangle_p$, so we describe the interferometry just after having applied the first beam splitter.\\

Tracking now the time-evolution of the density operator $[\![j_0(L),j_0(R)]\!]$ in the Heisenberg picture, we see that a phase rotation applied on the left mode now does manifest itself in the quantum observables of the Dirac field, by modifying the charge density operator! The reason is that the field transforms under the phase rotation $U_L^{(\phi)}$ at time $t$ as $b_L(t) + d_L^{\dagger}(t) \xrightarrow[\text{}]{U_L^{(\phi)}} b_L(t+dt) + d_L^{\dagger}(t+dt) = e^{i\phi}b_L(t) + e^{-i\phi}d_L^{\dagger}(t)$ and $b_R(t) + d_R^{\dagger}(t) \xrightarrow[\text{}]{U_L^{(\phi)}} b_R(t+dt) + d_R^{\dagger}(t+dt) = b_R(t) + d_R^{\dagger}(t)$. Then, after applying the phase rotation $U_L^{(\phi)}$, one obtains that the charge density of the Dirac field in the left mode is: 
\begin{equation}
    j_0(L) = -e\left(b_L^{\dagger}b_L +e^{-2i\phi} b_L^{\dagger}d_L^{\dagger} - e^{2i\phi} b_L d_L - d_L^{\dagger}d_L\right)\;,
\end{equation}
while the right mode  charge density stays unchanged.\\  

We see that the phase is present in the charge density operator of the left mode after applying the phase shift. So a perfectly valid observable, even under superselection rules, can keep track of the phase locally to each mode of the Dirac field. Of course, in parallel with what happened with the photon case, if the Heisenberg state consists of an electron superposed across the left and the right modes and no positrons, then the expected value of $j_0$ will still be phase independent-the phase is locally inaccessible. So, in the absence of superposed positrons, or another superposed electron that acts as a reference, the phase is at this point unobservable (all we can observe is whether the electron is in the mode $L$ or mode $R$). However, the local picture of quantum field theory, when considering the Dirac field in its entirety as a q-number, allows us to tell that the phase has been applied on one mode and not the other, by tracking the full description of the electron and positron field.\\ 

Moreover, in the Mach-Zehnder case, we can do so by tracking a physical observable, and not requiring the tracking of the Dirac field itself. In the Mach-Zehnder interferometer, the final beam-splitter mixes the left and the right modes in exactly the way to the phase to be observable in the output charge densities (or currents in general). After the final beam-splitter, the density operator in the left mode is
\begin{gather}
j_0 (L) = -\frac{e}{2} : \left(e^{-i\phi}b_L^\dagger + b_R^\dagger+ e^{i\phi}d_L + d_R\right) \left(e^{i\phi}b_L + b_R+ e^{-i\phi}d^{\dagger}_L + d^{\dagger}_R\right): = \nonumber \\ =-\frac{e}{2} \left[\left(e^{-i\phi}b_L^\dagger + b_R^\dagger\right) \left(e^{i\phi}b_L + b_R+ e^{-i\phi}d^{\dagger}_L + d^{\dagger}_R\right)-\left(e^{i\phi}b_L + b_R+ e^{-i\phi}d^{\dagger}_L + d^{\dagger}_R\right)\left(e^{i\phi}d_L + d_R\right)\right]
\end{gather}

Now, considering the expected value in the Heisenberg state $|\Psi\rangle_e = \frac{1}{\sqrt{2}} (|0_L 1_R\rangle_e + |1_L 0_R\rangle_e)|0_L 0_R\rangle_p$, we see that only four components of the density survive, i.e., $b^{\dagger}_L b_L+b^{\dagger}_R b_R + e^{-i\phi} b_L^{\dagger}b_R+e^{i\phi} b_R^{\dagger}b_L$. Two of these contain the phase information. The expected value in the state $|\Psi\rangle_e$ therefore is $-e\cos^2 \phi/2$, as expected, completely analogously to the bosonic case. As a side remark, we note that the operator $b_L^{\dagger}b_R + b_R^{\dagger}b_L$ quantifies coherence between the $L$ and $R$ modes (or what could be called a single particle entanglement between the two modes \cite{Vedral}). It is therefore not surprising that it emerges as the key observable in the Heisenberg treatment of interference. \\

\section{Discussion}

What have we achieved? We are able to describe the locality of spatial interference of both bosonic and fermionic fields in terms of the relevant local elements of reality, the q-numbered valued observables of the local fields pertaining to each mode, in the Heisenberg picture. The local elements of reality are in each case, the operators that are physically allowed observables, so considering the parity superselection rules for fermions. For bosons, it is sufficient to keep track of the photon field, while for fermions we need to use the current operator (or some other quadratic operator) of the full Dirac field.\\ 

Nevertheless, one could ask if this scheme can be applied to interacting fields. The answer is yes. In QED, for example, we have to combine the quantised electromagnetic field $A_\mu(x)$ with the Dirac field $\psi(x)$ through the charge current vector $j_\mu(x)=\frac{e}{2} [\bar{\psi},\gamma_\mu \psi(x)]$. In this case we have the quantum electrodynamics equations, that show the dependence of $A_\mu(x)$ with $A_{\mu}(x')$ and $j_\mu(x')$, and vice versa \cite{Kallen}. There we could follow the same strategy, tracking how the quantum observables $A_\mu(x), j_\mu(x)$ change. Since they only change through interactions local in position, all observed phases and phenomena can be explained in a non-action at a distance way using these observables as descriptors.\\

Using this scheme, it is possible to describe the Aharanov-Bohm effect in an entirely local way, \cite{AB} , by quantising the electromagnetic field and the Dirac field entirely and then using $j_\mu(x)$ and $A_\mu(x)$ as descriptors that track the local changes of the system. This can be done by generalising the Heisenberg picture treatment presented in \cite{MV}, using the picture of the full quantum field theory. We leave the details of this to a future work.\\   

We have previously questioned whether the fermionic phase due to anti-commutation is also acquired locally, \cite{MV-phase}. Here, instead, we embraced the anti-commutation of fermions and explored its consequences for a fully local description of quantum superpositions and interferences in quantum electrodynamics. The Heisenberg picture proved to be natural in this context. However, could, even more, be said? Could we argue that the Schr\"odinger picture does not even exist in some of the scenarios we have considered? The answer to this exciting question may be yes, since the operators describing interactions in quantum electrodynamics may not have a finite norm, i.e. they may not transform all state vectors with a finite norm into state vectors with the same property. Dirac provided an example of this when he emphasised that the Heisenberg picture helps us get rid of the deadwood in quantum electrodynamics arising due to the Schr\"odinger picture \cite{Dirac}. Our analysis adds another point in favour of the argument that the Heisenberg picture may be superior to the Schr\"odinger picture when it comes to showing that quantum theory and general relativity obey the same locality principle, albeit with different elements of reality: c-numbers for general relativity, q-numbers for quantum theory.  

{\sl Acknowledgments}: The Authors are grateful to David Deutsch for several fruitful discussions on the topic of locality. C.M. acknowledges funding from the Eutopia Foundation and the John Templeton Foundation. V.V. acknowledges funding from the National Research Foundation (Singapore), the Ministry of Education (Singapore) and Wolfson College, University of Oxford. This publication was made possible also through the support of the ID 61466 grant from the John Templeton Foundation, as part of the The Quantum Information Structure of Spacetime (QISS) Project (qiss.fr). The opinions expressed in this publication are those of the authors and do not necessarily reflect the views of the John Templeton Foundation. N.T.V states that the project that has generated these results has had the support of a scholarship from Fundaci\'on Bancaria La Caixa (ID 100010434), with code LCFBQEU1811650048.

\end{document}